\documentclass[11pt,twoside]{article}

\usepackage{asp2004}
\usepackage{epsf}
\usepackage{psfig}
\usepackage{lscape}
\usepackage{graphicx}

\begin{document}
\title{Discovery of the first very wide WD-LD binary system?}   %%% Fill in title
\author{Avril C. Day-Jones, David J. Pinfield, Hugh R.A. Jones, Tim R. Kendall, Ralf Napiwotzki and James S. Jenkins}   %%% Fill in author names
\affil{Centre for Astrophysics Research, University of Hertfordshire, College Lane, Hatfield, Herts, AL10 9AB}    %%% Fill in author affiliations

\begin{abstract} %%% Abstract to run on from here.
We report on our large scale search of 2MASS and SuperCOSMOS in the southern hemisphere for very widely separated white dwarf / L-dwarf binary systems and present our findings, including 8 widely separated candidate systems, and proper motion analysis confirming one of these as a widely separated white dwarf/ L-dwarf common proper motion binary candidate.
\end{abstract}

\section{Introduction}
In the last few years there have been many searches to find brown dwarf companions to white dwarfs. Despite these attempts only two brown dwarf / white dwarf binaries (with a brown dwarf spectral type later than M) have been confirmed. GD 165B (Zuckerman et al.~1992) and GD 1400 (Farihi et al.~2004), and both are known to be close systems, within separations of a few AU.  It is common place for brown dwarfs to exist in wide binaries to solar type and other low mass stars at separations of 1000-5000AU (Pinfield et al~ 2006; Gizis et al.~2001). However, when a star sheds its envelope as it moves into the white dwarf phase, we may expect a brown dwarf companion to migrate outwards (Burleigh, Clarke \& Hodgkin 2002) to separations of $\sim$4000-20,000AU.  Some may be broken apart quite rapidly by gravitational interactions with other stars, but some could survive.

The discovery of such binaries in which the white dwarf is high mass (and thus has a brief progenitor lifetime) will yield benchmark brown dwarfs where the cooling age of the white dwarf can, by association give a well constrained age for the brown dwarf. Cool dwarf spectra are highly dependent on these properties, but are notoriously difficult to model. Such benchmark brown dwarfs are thus vital to our interpretation of brown dwarf properties in general.

\section{Brown dwarf companions at wide separations}
Early analysis of data from 2MASS shows a high binary fraction of 18\% for wide companions around main sequence stars (Gizis et al.~2001). Our current work supports this high binary fraction and suggests a brown dwarf companion fraction of 34 $^{+9}_{-6}$ \%, suggesting that up to a third of all stars could have brown dwarf companions at wide separations (Pinfield et al.~2006). Thus far searches have revealed only 3  white dwarf/brown dwarf binaries, where the brown dwarf component is an L-dwarf and all of close separation, the widest of which is GD 165B which is at a separation of $\sim$150AU. So do such systems exist at much wider separations?

We have mined the 2MASS and SuperCOSMOS science archieves in the south, searching specfically for binary systems separated by up to 20,000AU. We Select suitable candidate white dwarfs using a series of colour, magnitude and reduced proper motion constraints established from known catalogued white dwarfs from SuperCOSMOS; and use colour and magnitude selection criteria (similar to those of Cruz et al.~2003) to select candidate L-dwarfs from 2MASS. 

\section {Results}
Our search revealed 8 candidate white dwarf/L-dwarf binaries, plotted in figures 1 and 2 show a colour-magnitude diagram of the L-dwarf candidates, a reduced proper motion of the white dwarf candidates and a plot of separation vs distance of the candidate binary pairs. Each of the candidate images were visually inspected in the R and I bands for counterparts and are consistent with having L-dwarf colours.  Proper motion follow up was done for 7 of the 8 candidates using second epoch observations from the Anglo Australian telescope and the William Herschal Telescope. We find 4 of the 8 are non common proper motion pairs and 2 are uncertain due to their small motion and high associated residuals.  However we find one of our candidates to be a common proper motion pair. Figure 3.~shows images of both the white dwarf and L-dwarf candidate components of the binary. The calculated proper motion for this system is.
\begin{itemize}
\item[]{\it L-Dwarf proper motion:} RA:~$-$130$\pm$30mas/yr,~DEC:~$-70\pm$20mas/yr. 
\item[]{\it White Dwarf proper motion:} RA:~$-$83$\pm$30mas/yr,~DEC:~$-$70$\pm$12mas/yr.
\end{itemize}
The measured proper motion in comparable to the proper motion of the white dwarf as recorded by SuperCOSMOS ~({\it pmRA:$-$78.5}mas/yr,~{\it pmDEC:$-$67.4}mas/yr). The L-dwarf colours ({\it J-K}=1.32, {\it J-H}=0.78) are consistent with an L2-3 type and the white dwarf colour ({\it B-R}=0.42) is consistent with a mid to hot white dwarf; suggesting that their distance is 35-44pc giving a binary separation of  3000-4000AU.

\section {Future Work}
We are awaiting follow up observations of the last candidate wide white dwarf / L-dwarf binary  and planare carrying out spectroscopic follow up and analysis of the common proper motion candidate to estimate Teff, gravity and determine an accurate spectral type for the candidate L-dwarf. We also plan to perform a similar search for such companions in UKIDSS (UKIRT Infrared Deep Sky Survey) and with 2MASS and SuperCOSMOS in the north when available. 

\begin{figure}
\plotfiddle{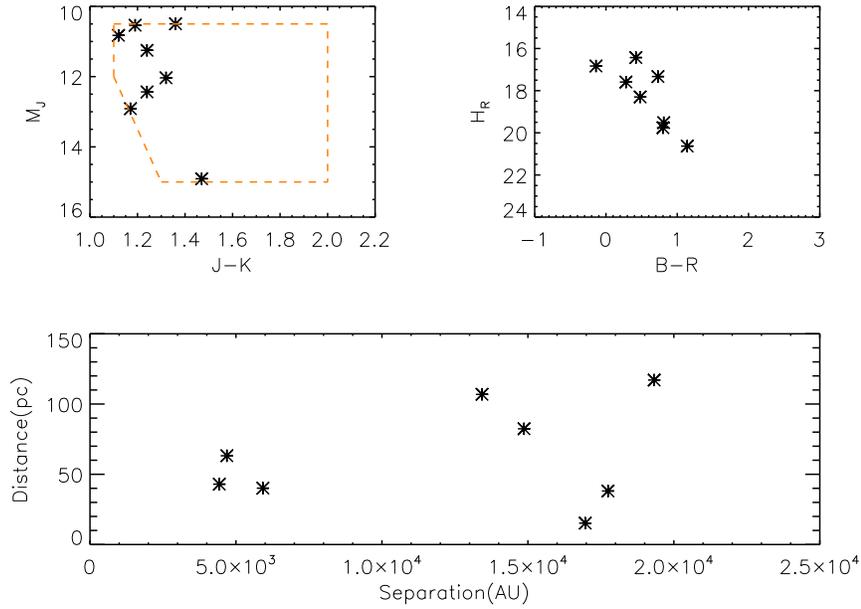}{3.0in}{90.}{49.}{49.}{180}{-25}
\caption{{\it Top Left}:~Colour-Mag Diagram of L-dwarf candidates (assumed to be at the same distance as the white dwarf),dashed box shows area occupied by known L-dwarfs. {\it Top Right}:~Reduced Proper Motion Diagram of white dwarf candidates. {\it Bottom}:~Distance vs Separation plot of binary candidates.}
\end{figure}

\begin{figure}
\plotfiddle{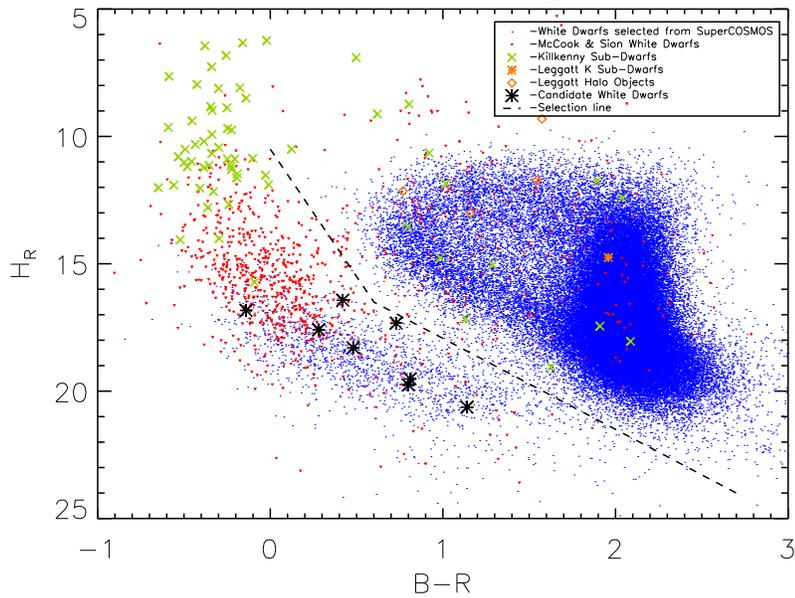}{3.0in}{90.}{50.}{50.}{180}{-25}
\caption{Enlarged Reduced Proper Motion Diagram of white dwarf candidates, overplotted with white dwarf selection area (to lower left of dashed line).}
\end{figure}

\clearpage

\begin{figure}[!h]
\plotfiddle{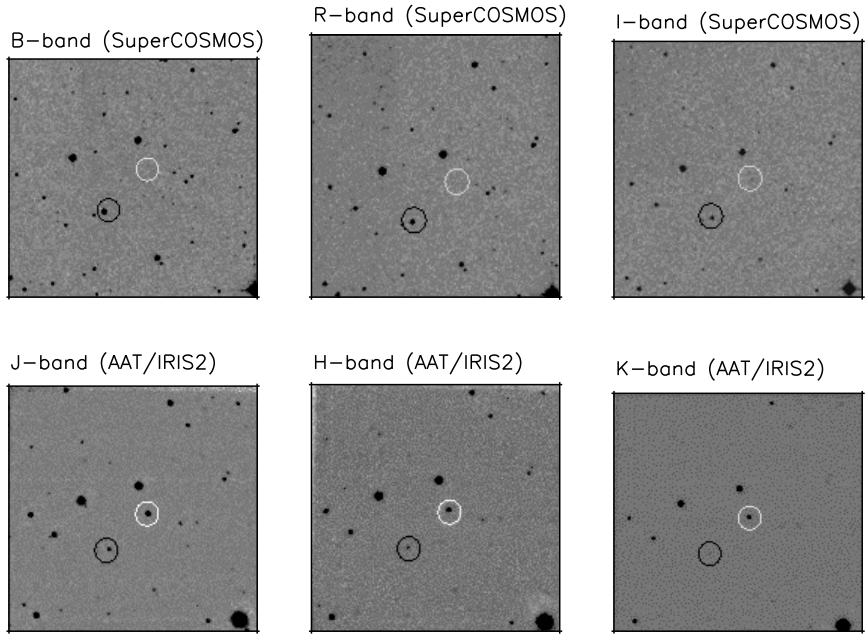}{3.0in}{90.}{50.}{50.}{190}{-45}
\caption{SuperCOSMOS and AAT/IRIS2 images of the white dwarf (white circles) and L-dwarf (black circles) common proper motion candidates. The white dwarf can be seen in the optical (B,R,I and just visible in the J band) and the L-dwarf can be seen in the Near Infrared (J,H and K bands). }
\end{figure}

%%% MAIN BODY OF TEXT GOES HERE. CONSULT "INSTRUCTIONS FOR AUTHORS USING
%%% LATEX2E MARKUP", SECTIONS 2.3-2.6 FOR HELP WITH EQUATIONS, FIGURES,
%%% AND TABLES.

%\section{}   %%% Top level section head (remove "%" symbol)
%\subsection{}   %%% Second level section head (remove "%" symbol)
%\subsubsection{}   %%% Lowest level section head (remove "%" symbol)
%\section*{}	%%% Unnumbered top level section head (remove "%" symbol)
%\subsection*{}   %%% Unnumbered second level section head (remove "%" symbol)

\acknowledgements %%% Text of acknowledgements runs on after this command.
ADJ acknowledges support from PPARC for this work. This research has made use of data from the SuperCOSMOS science archive and 2MASS (Two micron all sky survey).

%\begin{references}
%\reference bibliographic information
%Burleigh, M. R., Clarke, F. J., Hodgkin, S. T. 2002, MNRAS,
%\end{references}

%%% THE BIBLIOGRAPHY
%%%
%%% CONSULT SECTION 3 OF "INSTRUCTIONS FOR AUTHORS" FOR HOW TO USE NATBIB.
%%% AUTHORS ARE ENCOURAGED TO USE EITHER THE "THEBIBLIOGRAPY" ENVIRONMENT
%%% BY UNCOMMENTING (DELETING THE "%" SYMBOL) THE COMMANDS BELOW, OR BY
%%% USING THE BIBTEX ENVIRONMENT. TO FIND OUT WHICH IS APPLICABLE TO YOUR
%%% CONTRIBUTION, CONSULT THE VOLUME EDITORS FOR YOUR PROCEEDINGS.
%%%

\end{document}